\documentclass[aps,prd,superscriptaddress,nofootinbib,preprintnumbers,longbibliography,a4paper,11pt]{revtex4-2}
\usepackage{geometry}
\usepackage{mathtools}
\usepackage[utf8]{inputenc}
\usepackage[T1]{fontenc}
\usepackage{verbatim}
\usepackage{amsmath,amssymb,amsfonts}	
\usepackage{graphicx}
\usepackage{color}
\usepackage{xcolor}
\usepackage{physics}
\usepackage{booktabs}

\usepackage{appendix}
\usepackage{slashed}
\usepackage{tikz}
\usetikzlibrary{angles,quotes}
\usepackage[colorlinks=true,citecolor=blue,linkcolor=purple,urlcolor=purple]{hyperref}

\begin{document}
\preprint{}\preprint{CPPC-2023-05}
\heavyrulewidth=.08em
\lightrulewidth=.05em
\cmidrulewidth=.03em
\belowrulesep=.65ex
\belowbottomsep=0pt
\aboverulesep=.4ex
\abovetopsep=0pt
\cmidrulesep=\doublerulesep
\cmidrulekern=.5em
\defaultaddspace=.5em

\baselineskip 0.7cm

\bigskip

\title{Complementarity of $\mu$TRISTAN and Belle II in searches for charged-lepton flavour violation.}

\author{Gabriela Lichtenstein}
\email{g.lima_lichtenstein@unsw.edu.au}
\affiliation{Sydney Consortium for Particle Physics and Cosmology, School of Physics, The University of New South Wales,
Sydney, New South Wales 2052, Australia
}

\author{Michael A.~Schmidt}
\email{m.schmidt@unsw.edu.au}
\affiliation{Sydney Consortium for Particle Physics and Cosmology, School of Physics, The University of New South Wales,
Sydney, New South Wales 2052, Australia
}

\author{German Valencia}
\email{german.valencia@monash.edu
}
\affiliation{School of Physics, Monash University, Wellington Road, Clayton, Victoria 3800, Australia}

\author{Raymond R.~Volkas}
\email{raymondv@unimelb.edu.au}
\affiliation{ARC Centre of Excellence for Dark Matter Particle Physics, School of Physics, The University of Melbourne, Victoria 3010, Australia}

\begin{abstract}
We analyse the potential of the proposed $\mu^+ \mu^+$ and $\mu^+ e^-$ collider $\mu$TRISTAN to complement the searches for charged-lepton flavour-violation (CLFV) that can be carried out by Belle II. $\mu$TRISTAN offers the possibility of directly producing and studying new resonances that could mediate CLFV for a certain range of masses. In addition, we find that it can produce competitive bounds to those from Belle II for cases where the new resonance lies beyond direct reach. We illustrate these points with three $Z_3$ ``lepton triality'' models, where we also find an example that can only be probed by $\mu$TRISTAN. These three models feature doubly-charged scalars, denoted $k_{1,2,3}$ respectively, that induce both CLFV and flavour-conserving processes. Tree-level $k_1$ exchange induces the CLFV scattering process $\mu^+ e^- \to e^+ \tau^-$, while $k_2$ interactions induce $\mu^+ \mu^+ \to \tau^+ e^+$, $\mu^+ e^- \to \tau^+ \mu^-$ and make a non-SM contribution to the flavour-conserving scattering $\mu^+ \mu^+ \to \mu^+ \mu^+$. The $k_3$ model has a non-SM contribution to the flavour-conserving process $\mu^+ e^- \to \mu^+ e^-$. Other scattering processes involving $k_1$, $k_2$ or $k_3$ are not relevant for $\mu$TRISTAN and outside the scope of our analysis. We quantify the sensitivity of $\mu$TRISTAN for each of these processes. For the $k_1$ and $k_2$ cases we compare the $\mu$TRISTAN reach to the expected sensitivity of Belle II to the crossing symmetry related CLFV $\tau$ decays.

\end{abstract}

\maketitle

\newpage

\newpage

\section{Introduction}

Charged-lepton flavour violation (CLFV) is an important probe of many theories of physics beyond the standard model (BSM). Such processes must occur at some level because of the existence of lepton flavour violation in neutrino oscillations. However, to be experimentally accessible, there must be contributions to CLFV that are not proportional to the tiny neutrino mass eigenvalues. 

Motivated by the expected sensitivity of the Belle II experiment to CLFV decays of the $\tau$ lepton, models that extend the standard model (SM) by a doubly-charged scalar,  and whose Yukawa coupling structures are dictated by the discrete flavour symmetry of ``lepton triality''\footnote{The notion of lepton flavour triality has been first proposed in Ref.~\cite{Ma:2010gs}.}, were proposed in Ref.~\cite{Bigaran:2022giz}. Those models include both isosinglet and isotriplet extensions, but for the present paper we consider the doubly-charged isosinglet scalar cases only. The two isosinglet models of Ref.~\cite{Bigaran:2022giz} are briefly reviewed in Sec.~\ref{sec:triality} below and augmented by a related third model. In these models, the BSM decay processes are restricted by triality to pertain only to the $\tau$ lepton. In one class, the relevant decays are $\tau^\pm \to \mu^\pm \mu^\pm e^\mp$, while another class features $\tau^\pm \to e^\pm e^\pm \mu^\mp$. The third class does not have any kinematically-allowed CLFV lepton decays.

By crossing symmetry, flavour-changing scattering processes, in particular $\mu^+ \mu^+ \to \tau^+ e^+$, $\mu^+ e^- \to \tau^+ \mu^-$ and $\mu^+ e^- \to e^+ \tau^-$, are another feature of two of these models. The recent $\mu$TRISTAN proposal~\cite{Hamada:2022mua,Hamada:2022uyn}, which envisages both $\mu^+ \mu^+$ and $\mu^+ e^-$ colliders, is thus of direct relevance for testing these theories, and indeed for probing charged-lepton flavour physics in general~\cite{Fridell:2023gjx} in a way that is complementary to experiments such as Belle II. The purpose of this paper is to analyse the reach of these proposed colliders in the coupling constant and mass parameter space for these simple lepton triality BSM models. As well as the CLFV processes above, one of the models also has doubly-charged scalar exchange adding coherently with the SM contributions to the flavour-conserving scattering process $\mu^+ \mu^+ \to \mu^+ \mu^+$, while the new third model has its doubly-charged scalar contributing to $\mu^+ e^- \to \mu^+ e^-$. We explore the prospects for seeing these BSM flavour-conserving effects as well. 

The remainder of this paper is structured as follows. Section~\ref{sec:triality} reviews the isosinglet triality models of Ref.~\cite{Bigaran:2022giz} while Sec.~\ref{sec:constraints} summarises relevant experimental bounds. Section~\ref{sec:pheno-muTristan} provides the phenomenological analyses of the considered scattering processes and is the heart of the paper. We conclude in Sec.~\ref{sec:conc}.

\section{Lepton triality}
\label{sec:triality}

Lepton triality is a $Z_3$ flavour symmetry that acts only on leptons through the transformations
\begin{equation}
    L \to \omega^T L,\qquad e_R \to \omega^T e_R
\end{equation}
where $T=1,2,3$ for the first, second and third families, respectively, and $\omega = e^{2\pi i/3}$ is a cube root of unity. Thus electrons and electron neutrinos transform multiplicatively via $\omega$, muons and muon neutrinos transform via $\omega^2$, while taus and tau-neutrinos are triality singlets. This symmetry requires that the charged-lepton Yukawa and mass matrices are diagonal.\footnote{Neutrino mass generation is irrelevant for the purposes of this paper, so we refrain from reviewing the possibilities discussed in Ref.~\cite{Bigaran:2022giz}.} 
Lepton triality may emerge as an approximate symmetry in lepton flavour models with a discrete flavour symmetry which is broken to different subgroups for neutrinos and charged leptons, respectively, see e.g.~Refs.~\cite{Altarelli:2005yx,He:2006dk,deAdelhartToorop:2010jxh,deAdelhartToorop:2010nki,Cao:2011df,Holthausen:2012wz,Pascoli:2016wlt,Muramatsu:2016bda}.

We now introduce colourless, isosinglet, doubly-charged  scalars $k_1$, $k_2$ and $k_3$, with $k_1 \to \omega k_1$, $k_2\to \omega^2 k_2$ and $k_3 \to k_3$ under triality. We refer to these as the $T=1$, $T=2$ and $T=3$ cases, respectively, and they define the three models to be analysed in this paper. The $T=1$ and $T=2$ models were considered in Ref.~\cite{Bigaran:2022giz}, while the $T=3$ model is new. Although it may be phenomenologically interesting to introduce two or more scalars simultaneously, we consider one exotic per theory for the present analysis.

Triality dictates that 
$k_1$ is restricted to having the Yukawa interactions given by
\begin{equation}\label{eq:LagT1}
    \mathcal{L}_{k_1} = \frac12 \left( 2 f_1 \overline{(\tau_R)^c} \mu_{R} + f_2 \overline{(e_{R})^c} e_{R} \right) k_1 + \mathrm{h.c.} .
\end{equation}
while for $k_2$ the interactions are
\begin{equation}\label{eq:LagT2}
    \mathcal{L}_{k_2} = \frac12 \left( 2 g_1 \overline{(\tau_{R})^c} e_{R} + g_2 \overline{(\mu_{R})^c} \mu_{R} \right) k_2 + \mathrm{h.c.}
\end{equation}
and for $k_3$ they are
\begin{equation}\label{eq:LagT3}
    \mathcal{L}_{k_3} = \frac12 \left( 2 h_1 \overline{(\mu_R)^c} e_{R} + h_2 \overline{(\tau_{R})^c} \tau_{R} \right) k_3 + \mathrm{h.c.} ,
\end{equation}
where $f_{1,2}$, $g_{1,2}$ and $h_{1,2}$ are coupling constants which can be chosen real and non-negative without loss of generality. The masses of the exotic scalars are denoted by $m_{k_1}$, $m_{k_2}$ and $m_{k_3}$. The $k_{1,2}$ models feature exotic $\tau$ decays, while $k_3$ contributes to scattering processes.

\begin{table}[t]
    \begin{centering}
    \begin{minipage}{0.5\linewidth}
    \begin{ruledtabular}
    \begin{tabular}{ccc}
    Model & Process & Lepton Collider \\\hline
    T=1 & $\mu^+ e^- \rightarrow e^+ \tau^- $ & $\mu$TRISTAN \\
    T=1 & $e^+ e^- \rightarrow e^+ e^-$ & $e^+ e^-$ \\ 
    T=1 & $e^- e^- \rightarrow e^- e^- $ & - \\
     T=1 & $e^- e^- \rightarrow \tau^- \mu^- $ & -  \\
     
    T=2 & $\mu^+ \mu^+ \rightarrow \tau^+ e^+ $ & $\mu$TRISTAN \\
    T=2 & $\mu^+ \mu^+ \rightarrow \mu^+ \mu^+ $ & $\mu$TRISTAN \\
    T=2 & $\mu^+ e^- \rightarrow \tau^+ \mu^- $ &
    $\mu$TRISTAN \\
     T=2 & $\mu^+ \mu^- \rightarrow \mu^+ \mu^- $ & $\mu^+\mu^-$ \\
     T=3 & $\mu^+ e^- \to \mu^+ e^-$ & $\mu$TRISTAN\\
     T=3 & $\mu^+ e^+ \to \tau^+ \tau^+$ & -\\
    \end{tabular}
    \end{ruledtabular}
    \end{minipage}
    \end{centering}
    \caption{List of all tree-level $2\to2$ charged-lepton scattering processes with $e^\pm$ and $\mu^\pm$ in the initial state mediated by the doubly charged scalars in the isosinglet lepton-triality models. The proposed $\mu$TRISTAN collider can probe half of these processes, three of them violating flavour and two conserving flavour. Electron-positron and muon-antimuon colliders can probe two additional flavour-conserving scattering processes, an antimuon-positron collider could probe the process with two tau leptons, while an electron-electron collider would be needed to cover the remaining two.}
    \label{tab:list}
\end{table}

We have listed all possible tree-level $2\to 2$ charged-lepton scattering processes (up to particle-antiparticle swaps) that can be searched for using various lepton colliders in Table~\ref{tab:list}. The proposed $\mu$TRISTAN collider, in both its $\mu^+ \mu^+$ and $\mu^+ e^-$ configurations, would be able to probe the CLFV scattering processes of the $T=1,2$ triality models. The same-sign muon collider would also be able to search for a resonant $s$-channel $k_2$ contribution to $\mu^+ \mu^+ \to \mu^+ \mu^+$, while the $\mu^+ e^-$ collider could search for $u$-channel $k_3$ contribution to $\mu^+ e^- \to \mu^+ e^-$. Electron-positron colliders could also search for $u$-channel $k_1$ contributions to $e^+ e^- \to e^+ e^-$, while an opposite-sign muon collider would be sensitive to $u$-channel $k_2$ exchange affecting $\mu^+ \mu^- \to \mu^+ \mu^-$. 
The remaining three processes listed in Table~\ref{tab:list} require either an electron-electron or an antimuon-positron collider, for which there are no current proposals.

\section{Existing constraints}
\label{sec:constraints}

We now briefly quote some existing experimental constraints. ATLAS \cite{ATLAS:2017xqs} searches on pair production of doubly-charged scalar singlets decaying to same-sign charged leptons $l^\pm l^\pm$ produce direct bounds on the scalar masses. 
To maximise the Yukawa couplings, we consider the case where the doubly-charged scalar has a $50\%$ branching ratio to electrons, muons or an electron-muon pair and the remaining $50\%$ to $\tau \mu$, $\tau$e, or $\tau\tau$ respectively. Therefore, considering the detector would miss all events with $\tau$ leptons, the lower bounds on the doubly-charged scalars are:
\begin{equation}
    m_{k_i} > 0.6 \text{ TeV }\quad \text{for}\quad i=1,2,3.
    \label{eq:mkConstraints}
\end{equation}
Alternatively, for a $100\%$  branching ratio to electrons, muons or an electron-muon pair, the lower bounds are:
\begin{equation}
    m_{k_1} > 0.6 \text{ TeV, } \ \ 
    m_{k_2} > 0.7 \text{ TeV, } \ \ 
    m_{k_3} > 0.8 \text{ TeV. }
\end{equation}

Recent direct searches at the LHC~\cite{ATLAS:2022pbd} suggest an improvement of these constraints. They present an upper bound on the doubly charged scalar production cross-section derived from more data than the previous result. However, the bounds on the singlet mass are obtained considering universal lepton couplings. They find that the limit is improved to $m_{k_i} > 0.9$ TeV for the case of the doubly-charged scalar in the Zee-Babu model~\cite{Zee:1985id,Babu:1988ki}. This limit should be comparable to those for the triality models when the doubly-charged scalar decays to same-sign leptons with equal probability for different flavours and thus an improvement over the constraint quoted in Eq.~\ref{eq:mkConstraints}.

Reference~\cite{Bigaran:2022giz} contains an analysis of the existing Belle constraints on the CLFV decay modes of the $\tau$, which produce the parameter bounds
\begin{equation}
    \sqrt{f_1 f_2} \lesssim 0.17\, \frac{m_{k_1}}{\mathrm{TeV}}\quad \text{and}\quad
    \sqrt{g_1 g_2} \lesssim 0.17\, \frac{m_{k_2}}{\mathrm{TeV}}.
    \label{Eq:UpperBound}
\end{equation}
Reference~\cite{Bigaran:2022giz} also determined that the eventual Belle II sensitivity reach for the scalar masses is given by $m_{k_1} \lesssim 61$ TeV and $m_{k_2} \lesssim 59$ TeV, assuming 
they record $50\,\text{ab}^{-1}$ 
of data. This can also be expressed as upper bounds on the parameters
\begin{equation}
    \sqrt{f_1 f_2} \lesssim 0.06\, \frac{m_{k_1}}{\mathrm{TeV}}\quad \text{and}\quad
    \sqrt{g_1 g_2} \lesssim 0.06\, \frac{m_{k_2}}{\mathrm{TeV}}.
\end{equation}%
The scalar $k_3$ cannot be probed by tau decays at tree level. Further constraints are set by the DELPHI analysis of contact interactions~\cite{DELPHI:2005wxt}, which can be translated to a bound on the Yukawa couplings~\cite{Li:2018cod,Li:2019xvv},
\begin{equation}
\begin{aligned}
    f_2 &\lesssim 1.4 \, \frac{m_{k_1}}{\mathrm{TeV}}\;, &
    g_1 &\lesssim 0.66 \, \frac{m_{k_2}}{\mathrm{TeV}}\;, &
    \mathrm{and}\;\;h_1 &\lesssim 0.72 \, \frac{m_{k_3}}{\mathrm{TeV}}\;.
\end{aligned}
\label{eq:DELPHI}
\end{equation}
Constraints may also be obtained using other $e^+e^-$ scattering data, such as from SuperKEKB. Finally, the doubly-charged scalars contribute to leptonic anomalous magnetic moments. The contribution is always positive and thus may be able to alleviate the tension in the muon anomalous magnetic moment for large Yukawa couplings.

\section{Phenomenology at $\mu$TRISTAN}
\label{sec:pheno-muTristan}

The $\mu$TRISTAN proposal consists of a high energy lepton collider using the ultra-cold antimuon technology developed for the $g-2$ experiment at J-PARC~\cite{Hamada:2022mua}. There are two possible modes of operation. The first mode features a $\mu^+$ beam colliding with an $e^-$ beam produced at TRISTAN in a storage ring of circumference 3 km.  The $\mu^+ e^-$ configuration would be asymmetric in energy, typically with a centre-of-mass energy of 346 GeV from a 30 GeV and 1 TeV electron and antimuon beam, respectively.  We also consider the centre-of-mass energy of 775 GeV from a 50 GeV electron and 3 TeV antimuon beam. The second mode is a $\mu^+ \mu^+$ collider with both beams having energies of 1 TeV, thus producing a 2 TeV centre-of-mass energy. Following Ref.~\cite{Hamada:2022mua}, we also consider the $\sqrt{s} = 1,\, 3,\, 4$ TeV cases. The luminosity is estimated to be ${\cal L} = 4.6 \times 10^{33}$ cm$^{-2}$ s$^{-1}$  for the $\mu^+ e^-$ configuration and ${\cal L} = 5.7 \times 10^{32}$ cm$^{-2}$ s$^{-1}$  for the $\mu^+ \mu^+$ collider, which is reduced to about $70\%$ of these values when various efficiencies are taken into account -- see Sec.~2.4 in Ref.~\cite{Hamada:2022mua}. Changing units, the final luminosities we use are thus $100\ \text{fb}^{-1} \ \text{year}^{-1}$ for $\mu^+ e^-$ and $12\ \text{fb}^{-1}\ \text{year}^{-1}$ for $\mu^+ \mu^+$.
To obtain the following results we have implemented the model in  FeynRules \cite{Christensen:2009jx,Alloul:2013bka} to generate  FeynArts \cite{Hahn:2000kx} output and simplification is assisted with FeynCalc \cite{Mertig:1990an,Shtabovenko:2016sxi}. Processes with SM interference were also cross-checked with Madgraph \cite{Alwall:2014hca}.

\subsection{Lepton flavour-violating signals}

As summarised in Table~\ref{tab:list}, the triality models produce three CLFV $2 \to 2$ scattering channels at tree level that can be searched for using $\mu$TRISTAN.  One channel can be probed at the $\mu^+ \mu^+$ collider, and the other two through the $\mu^+ e^-$ collider:
\begin{equation}
    \mu^+ \mu^+ \rightarrow \tau^+ e^+ \ (T=2),\qquad
    \mu^+ e^- \rightarrow \tau^+ \mu^- \ (T=2), \qquad
    \mu^+ e^- \rightarrow e^+ \tau^- \ (T=1),
\end{equation}
which are related by crossing symmetry to the CLFV tau decays mentioned above.
We now calculate the event rate for each process using the luminosities quoted above, and explore the dependence on centre-of-mass energy, the masses $m_{k_i}$ and the coupling constants $f_{1,2}$ and $g_{1,2}$.

\subsubsection{Same-sign antimuon collider}

\begin{figure}[t]
         \centering
           \includegraphics[width=0.6\textwidth]{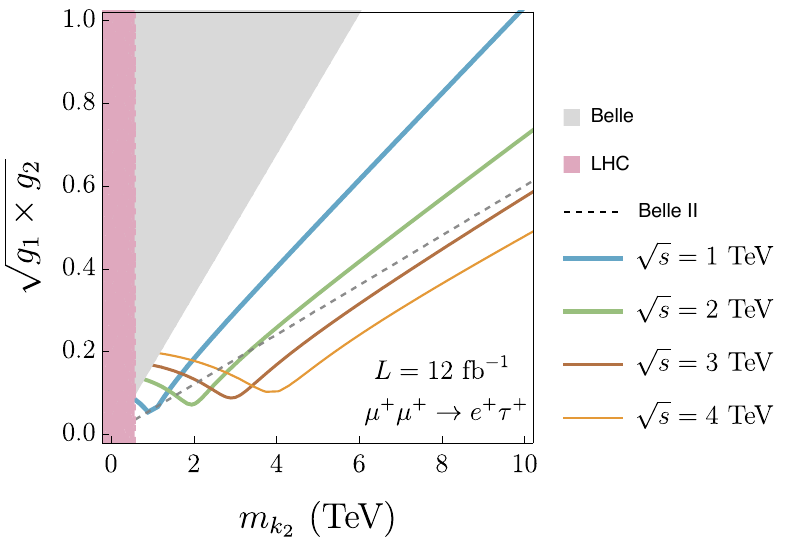}
      
\caption{ The coloured contours show the 90\% C.L.\ upper bound from $\mu^+ \mu^+ \to e^+ \tau^+$ on $\sqrt{g_1g_2}$ as a function of the scalar mass for $\sqrt{s} =$ 1, 2, 3, 4 TeV assuming no signal or background events are seen after $1$ year of running. The effect of the $s$-channel resonance is clearly seen. The current Belle exclusion region is shown in grey, while the dashed line shows the eventual Belle II sensitivity with 50 ab$^{-1}$ of data.
The magenta band is the region excluded by direct searches at the LHC, which constrains $m_{k_2} >  0.6 $ TeV as per Eq.~\eqref{eq:mkConstraints}. 
}
\label{f:90CLmumu}
     \end{figure}

The differential cross section for $\mu^+ \mu^+ \rightarrow \tau^+ e^+ $ in the $T=2$ model is given by
\begin{equation}
    \label{eq:crossmumu}
    \frac{d\sigma}{d\Omega} = \frac{g_1^2 g_2^2}{256 \pi^2} \frac{s}{(s-m_{k_2}^2)^2 + m_{k_2}^2 \Gamma^2_{k_2}} \qquad
    \mathrm{with}\qquad
    \Gamma_{k_2}   = \frac{g_1^2 m_{k_2}}{16 \pi} + \frac{g_2^2 m_{k_2}}{32 \pi}
\end{equation}
where $\Gamma_{k_2}$ is the total width of the $k_2$ resonance and the relatively small lepton masses have been neglected. Taking into account the angular detector acceptance $16^\circ < \theta < 164^\circ$~\cite{Hamada:2022uyn} and adopting a luminosity of ${\cal L} = 12\ \mathrm{fb}^{-1}\ \mathrm{year}^{-1}$, we determine the 90\% C.L.\ regions of parameter space after 1 year of running, as shown in Fig.~\ref{f:90CLmumu}. The blue, green, brown and orange lines depict the 90\% C.L. bound that would be obtained if no events were detected for different centre-of-mass energies $\sqrt{s}=1,2,3,4$ TeV. The regions above these lines produce event rates that are distinguishable from zero at $\ge 90$ \% C.L. The region excluded by current LHC data is indicated by magenta on the left. The region excluded by Belle~\cite{Hayasaka:2010np} from their bound on BR($\tau^+ \to \mu^+\mu^+ e^-)$ is given by the grey-shaded region. The eventual Belle II reach~\cite{Banerjee:2022xuw} with 50 ab$^{-1}$ of data is indicated by the grey dashed line. 

 Note that even though Eq.~(\ref{eq:crossmumu}) depends on $g_1$ and $g_2$ separately through the decay width $\Gamma_{k_2}$, we have quoted the bound as a function of $\sqrt{g_1 g_2}$. To define the boundary of the sensitivity region, the lowest cross-section is adopted, which occurs when the width is the largest it can be with perturbative coupling constants $g_{1,2} < \sqrt{4 \pi}$. This effectively renders the 90 \% C.L.\ line as depending only on $\sqrt{g_1 g_2}$. In the effective field theory limit, it results in
\begin{equation}
\sqrt{g_1g_2}\lesssim 0.15\left(\frac{N}{L s}\right)^{1/4}\frac{m_{k_2}}{\rm TeV}
\;.
\end{equation}

\subsubsection{Antimuon-electron collider}

We now turn to the processes $\mu^+ e^- \rightarrow e^+ \tau^- $ ($T=1$) and $\mu^+ e^- \rightarrow \tau^+ \mu^- $ ($T=2$) which are mediated in the $u$-channel by $k_1$ and $k_2$, respectively. Note that, with lepton masses neglected, the $T=2$ cross-section may be obtained from the $T=1$ cross-section simply by the replacements $f_{1,2} \to g_{1,2}$ and $m_{k_1} \to m_{k_2}$. In computing the event rate, account must be taken of the angular acceptance, since there will be a shield around the antimuon beam.

\begin{figure}[t]
      \begin{tikzpicture}[scale=1.5]

\node (ie) at (4,0) {$e^-$};
\node (imu) at (8,0) {$\mu^+$};
\node (c) at (6,0) {};
\node (fe) at (6.8,1.5) {$e^+$};
\node (ftau) at (4.5,-1) {$\tau^-$};

 \draw[black, thick, -stealth] (ie) -- (c);
\draw[black, thick, -stealth] (imu) -- (c);

\draw[black, thick, -stealth] (c) -- (fe);
\draw[black, thick, -stealth] (c) -- (ftau);

\draw[blue, dashed] (6,0) -- (8,.5);
\draw[blue, dashed] (6,0) -- (8,-.5);
\filldraw[blue] (9,1) circle (0pt) node[anchor=north]{shield};

\pic[draw,<->,angle radius=8mm, angle eccentricity=1.5, "$\phi_{ee'}$"] {angle = imu--c--fe};
\pic[draw,<->,angle radius=8mm, angle eccentricity=1.5, "$\phi_{\mu\tau'}$"] {angle = ie--c--ftau};

\end{tikzpicture}
  \caption{A sketch of the kinematics of $\mu^+ e^- \to e^+ \tau^-$ scattering in the laboratory frame. The shield around the $\mu^+$ beam is indicated. The angle between the 3-momentum of a final state lepton with the antimuon beam has to lie inside $[15.4^\circ,178^\circ]$.}
  \label{fig:kinematics}
  \end{figure}
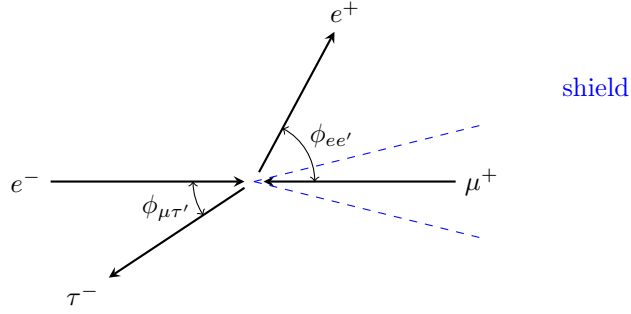

The $\mu^+ e^- \rightarrow e^+ \tau^-$ differential cross-section is given by:
\begin{equation}
    \frac{d\sigma}{du} = \frac{f_1^2 f_2^2}{64 \pi s^2} \frac{u^2}{(u-m_{k_1}^2)^2} ,
\end{equation}
and the total cross-section is
\begin{equation}
   \sigma = \frac{f_1^2 f_2^2}{64 \pi s^2} \left[u_+-u_-+\frac{m_{k_1}^4}{m_{k_1}^2-u_+} - \frac{m_{k_1}^4}{m_{k_1}^2-u_-}+2m_{k_1}^2 \ln\frac{m_{k_1}^2-u_+}{m_{k_1}^2-u_-}\right] 
\end{equation}
where  $u_-\leq u \leq u_+$. The integration limits $u_\pm$ are defined by the kinematics in the laboratory frame, as depicted in Fig.~\ref{fig:kinematics}. We are using $e$ to denote the initial $e^-$ state and $e'$ to denote the final $e^+$ state, with $\phi_{ij}$ being the angle between particles $i$ and $j$ in the laboratory frame. Therefore, the Mandelstam variable $u$ is
\begin{align}
    u&= -2 E_e E_e^\prime (1 - \cos\phi_{ee'}) = -2 E_\mu E_\tau^\prime (1-\cos\phi_{\mu\tau'}),
    \label{eq:intLimLabframe1}
\end{align}
where $\vec{p}_e \cdot {\vec{p}\, }'_{e} = E_e E_e'\cos\phi_{ee'}$ and similarly for $\phi_{\mu\tau'}$. Using energy-momentum conservation, we find the final state energies in terms of the initial state energies scattering angles,
\begin{align}
    E_e^\prime & = \frac{2 E_e E_\mu}{E_e (1-\cos\phi_{e e'}) + E_\mu (1+\cos\phi_{ee'})}, 
    &
    E_\tau^\prime & = \frac{2 E_e E_\mu}{E_e(1+\cos\phi_{\mu\tau'}) + E_\mu (1-\cos\phi_{\mu\tau'})}\;,
    \label{eq:intLimLabframe2}
\end{align}
which recovers what we expect in the extreme cases. For no scattering, $\cos\phi_{ee'}=\cos\phi_{\mu\tau'}=1$, there is no 4-momentum transfer ($u=0$) and thus the energies of the final state particles are the same as for the initial state particles. Similarly for $\cos\phi_{\mu\tau'}=-1$, where the incoming particles reverse direction, momentum conservation demands maximum 4-momentum transfer $u=-s$ and thus the energies of the particles are just swapped, i.e.\ the energy of the final state positron $E_e'$ equals $E_\mu$, and similarly $E_\tau' = E_e$.

The shield around the antimuon beam limits the detector acceptance: The angle between the 3-momentum of a final state lepton with the muon beam has to lie inside $[15.4^\circ,178^\circ]$. (Note that the electron beam also defines a small angular region outside the acceptance region.)  This can be re-expressed in terms of a constraint on the Mandelstam variable $u$. For $E_e=30$ GeV and $E_\mu=1$ TeV, we find $u\in [-1.188\times 10^{5},-1206]\  \mathrm{GeV}^2$.

\begin{figure}[t]
         \centering
          
    \includegraphics[width=0.6\textwidth]{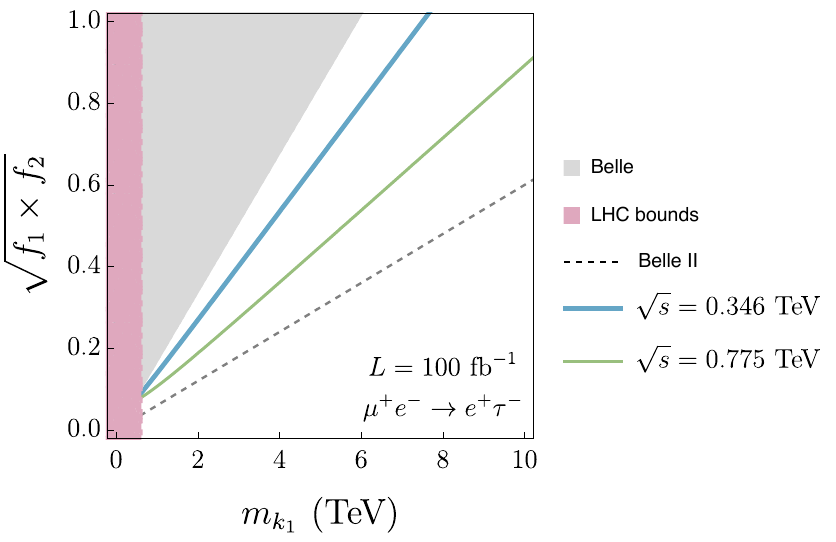}
      
\caption{The coloured contours show the 90\% C.L.\ upper bounds from $\mu^+ e^- \to e^+ \tau^-$ on $\sqrt{f_1f_2}$ as a function of the scalar mass for $\sqrt{s} =$ 0.346, 0.775 TeV, with asymmetric beam energies of $E_\mu =$ 1, 3 TeV and $E_e =$ 30, 50 GeV respectively. We assume no signal or background events are seen after $1$ year of running.   The current Belle exclusion region is shown in grey, while the dashed line shows the eventual Belle II sensitivity with 50 ab$^{-1}$ of data.
The magenta band is the exclusion region from direct searches at the LHC, which constrains $m_{k_1} >  0.6 $ TeV as per Eq.~\eqref{eq:mkConstraints}.
}
\label{f:90CLmue}
     \end{figure}

The parameter space in Fig.~\ref{f:90CLmue} shows the $\mu^+ e^-$ collider 90\% C.L.\ sensitivity curves in the ($\sqrt{f_1 \, f_2}$, $m_{k_1}$) plane after $1$ year of data taking. The curves are well approximated by the effective field theory (EFT) limit, which we compute to be $\sqrt{f_1f_2}\lesssim 0.13\frac{m_{k_1}}{\rm TeV}$. The grey and magenta regions have the same meaning as for Fig.~\ref{f:90CLmumu}. As stated earlier, the result  for the channel $\mu^+ e^- \rightarrow \tau^+ \mu^-$ in the $T=2$ model can be obtained by the replacements $f_{1,2} \to g_{1,2}$ and $m_{k_1} \to m_{k_2}$.

\subsubsection{Tau Identification}

Due to its short lifetime of $\tau_\tau = 2.903 \times 10^{-13}\ \text{s}$ , tau identification is usually performed using its decay products.
However, in a high energy lepton collider such as $\mu$TRISTAN, the tau will be boosted resulting in a displaced vertex signal. For instance, considering all leptons to be massless in the process $\mu^+ e^- \rightarrow e^+ \tau^-$ , the absolute value of the final-state tau momentum is given by $p_\tau = \frac{ \sqrt{s}}{2} \simeq E_\tau$. The corresponding decay length of the tau is then $\beta\gamma_\tau\tau_\tau$, where $\gamma_\tau=\frac{E_\tau}{m_\tau}$. We have estimated these displaced vertices to occur approximately $5$ cm from the collision point for $\sqrt{s} = 2$ TeV at the $\mu^+\mu^+$ collider.

\subsection{Doubly-charged scalar resonance effects}

We now turn to some lepton-flavour conserving processes that are altered from their SM expectations through doubly-charged scalar exchange effects. 

\subsubsection{$\mu^+ \mu^+ \rightarrow \mu^+ \mu^+ $ in the $T=2$ model}

\begin{figure}[t]
    \centering
    \includegraphics[width=0.7\textwidth]{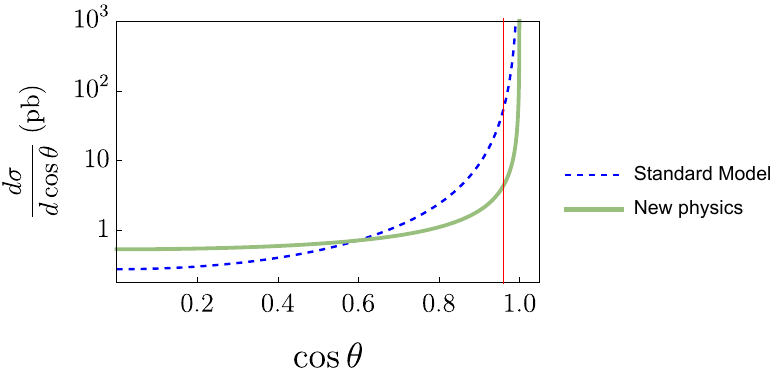}
    \caption{The differential cross-section for $\mu^+ \mu^+ \to \mu^+ \mu^+ $ in terms of the scattering angle $\cos \theta$ for the benchmark choices $g_1 = g_2 = 1$, $m_{k_2} = 1 $ TeV and $\sqrt{s} = 2 $ TeV. The blue dashed line corresponds to the SM cross-section, while the green is the difference between the full differential cross-section and the SM cross-section. The red vertical line indicates the boundary of the angular acceptance region.}
    \label{fig:difcs}
\end{figure}

 In this case, a tree-level $s$-channel $k_2$ exchange diagram interferes with the SM electroweak amplitude. A routine calculation produces the differential cross-section
 \begin{align}
     \frac{d \sigma}{d\cos\theta} &= \frac{\pi \alpha^2}{4 s^4_W c^4_W} \frac{1}{s} 
     \frac{1}{\sin^4\theta} \left( 1 + 17 s_W^4 + s_W^4 \cos^2\theta (6 + \cos^2\theta) \right)\nonumber\\
     &+ \frac{\alpha g_2^2}{4 c_W^2} \frac{1}{\sin^2\theta} \frac{s - m^2_{k_2}}{(s - m^2_{k_2})^2 + m^2_{k_2} \Gamma_{k_2}^2} + \frac{g_2^4}{256 \pi} \frac{s}{(s - m_{k_2}^2)^2 + m^2_{k_2} \Gamma_{k_2}^2}
 \end{align}
where $\alpha$ is the fine-structure constant, $s_W$ and $c_W$ are the sine and cosine of the weak mixing angle $\theta_W$, and the lepton and $Z$-boson masses were neglected. To obtain the following results we used $\alpha \simeq 1/128$ and $\sin^2 \theta_W \simeq 0.23$. The right-hand side of the above equation is the sum of the SM term, the interference term, and the direct $k_2$ term, reading from left to right. The difference between the full differential cross-section and the SM differential cross-section is plotted in green in Fig.~\ref{fig:difcs} for $m_{k_2} = 1$ TeV, $g_1 = g_2 = 1$ and $\sqrt{s} = 2$ TeV, and compared with the SM cross-section. The figure shows the familiar forward divergence of M\"oller scattering for the SM term, $\sim\sin^{-4}\theta$, as well as the one in the interference term which behaves as $\sim\sin^{-2}\theta$. The figure demonstrates the effect of the angular detector acceptance $16^\circ < \theta < 164^\circ$, represented by the vertical red line, which serves as a cut to suppress the SM background. The angular cut can thus be used to optimise the signal significance, although the precise value where this happens depends on the parameters of the model.

\begin{figure}[t]
       \includegraphics[width=0.8\textwidth]{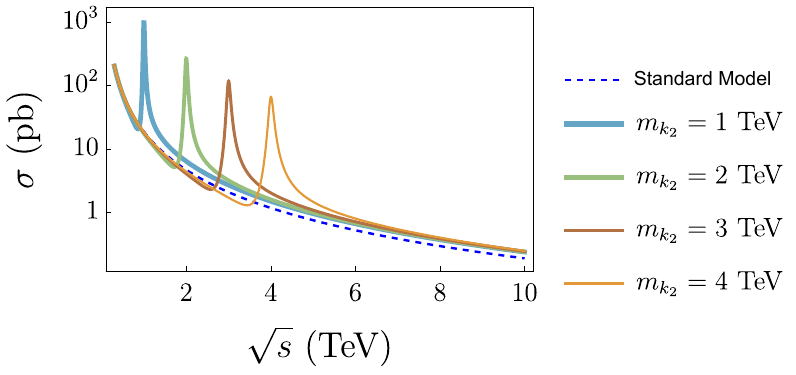}
         \caption{Total cross-sections for $\mu^+ \mu^+ \to \mu^+ \mu^+$ as functions of the centre of mass energy $\sqrt{s}$. The blue dashed curve is the purely SM cross-section, while the other full lines include the $k_2$ contribution with $g_1=g_2 = 1$ for $m_{k_2} = 1, 2, 3, 4$ TeV.}
        \label{fig:cross3}
    \end{figure}

The total cross-section, integrated over the angular detector acceptance region is given by
\begin{align}
    \sigma\simeq&\left(\frac{18.8}{s}-3.9\, g_2^2\, \frac{m^2_{k_2}-s}{(s-m^2_{k_2})^2 + m^2_{k_2} \Gamma_{k_2}^2}+0.9\, g_2^4\, \frac{s}{(s - m^2_{k_2})^2 + m^2_{k_2} \Gamma_{k_2}^2}\right){\rm~ pb} 
    \label{eq:numsigma}
\end{align}
Figure~\ref{fig:cross3} shows the full total cross-section for $m_{k_2} = 1, 2, 3, 4$ TeV and $g_2 = 1$, together with the purely SM total cross-section, as functions of $\sqrt{s}$. The resonances and the effects of both destructive and constructive interference are clearly visible.

We define the signal significance $S$ as the ratio of the number of events due to $k_2$ exchange (including the interference term) to the square root of the number of SM background events. Thus
\begin{align}
    S=\frac{|\sigma-\sigma_{SM}|}{\sqrt{\sigma_{SM}}}\sqrt{L}
\end{align}
where $\sigma$ is the full total cross-section, $\sigma_{SM}$ is the SM cross-section and $L$ is the integrated luminosity. We estimate the 90\% confidence level limit through $S=1.64$. In the EFT ($s \ll m_k^2$) limit we obtain an upper limit on the Yukawa coupling
\begin{align}
    g_2&\lesssim 0.18 \left(\frac{S^2}{{L}s}\right)^{1/4}m_{k_2},
\end{align}
with ${L}$ in fb$^{-1}$ and $\sqrt{s},~m_k$ in TeV, meaning that if this inequality is violated  then the non-SM effects are detectable at greater than 90\% C.L. In this EFT limit the interference term dominates and is negative, so that a $k_2$-induced signal corresponds to a deficit of events with respect to those expected in the SM. For $\sqrt{s}\sim{\cal O}(\rm TeV)$ as considered above, observing the fluctuations corresponding to the 90\% C.L.\ limit implies knowledge of the SM cross-section at the tenth of a percent level.

\begin{figure}[t]
    \centering
    \includegraphics[width=0.6\textwidth]{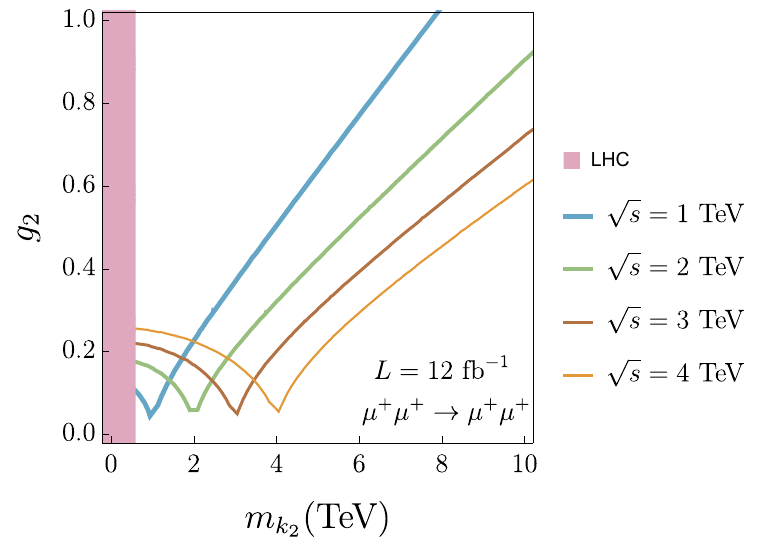}
    \caption{The $90\%$ C.L.\ contours in the parameter space $(g_2, m_{k_2})$ in the $T=2$ for the flavour-conserving scattering $\mu^+ \mu^+ \to \mu^+ \mu^+$ with $g_1 = 1$. The regions above the curves produce signals that are distinguishable from the SM expectation at $\ge 90$\% C.L.\ after $1$ year of data taking. The effect of the $s$-channel resonance is clearly seen. The limit is based on a deficit (excess) of events with respect to the SM below (above) the resonance as can be seen in  Figure~\ref{fig:cross3}. The magenta region is excluded by direct searches at the LHC.} 
    \label{fig:sminterference}
\end{figure}

Going beyond the EFT limit, the full 90\% C.L.\ contours are shown in Fig.~\ref{fig:sminterference} for $g_1 = 1$ and $L = 12\ \text{fb}^{-1}$, the latter being the expected integrated luminosity for a year of data taking. The regions above the contours produce $k_2$-induced effects that are distinguishable from the SM background expectation at greater than 90\% C.L.\ Depending on the regime, the induced effects either decrease or increase the number of events relative to the SM expectation.

\subsubsection{$\mu^+ e^- \to \mu^+ e^-$ in the $T=3$ model}

\begin{figure}
    \centering
    \includegraphics[width=0.6\textwidth]{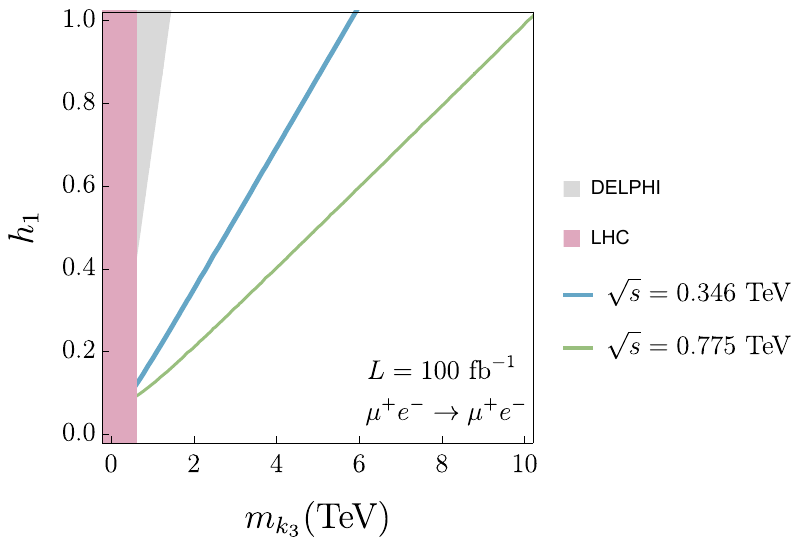 }
    \caption{The $90\%$ C.L.\ contours in the parameter space $(h_1, m_{k_3})$ of the $T=3$ model for the flavour-conserving scattering $\mu^+ e^- \to \mu^+ e^-$, which is independent of $h_2$. The regions above the curves produce signals that are distinguishable from the SM expectation at $\ge 90$\% C.L. after $1$ year of data taking. The grey region is excluded by the DELPHI limit quoted in Eq.~(\ref{eq:DELPHI}), and the magenta region is excluded by direct searches at the LHC.
    }
    \label{fig:T3paramsp}
\end{figure}

The $T=3$ model sees $\mu^+ e^- \to \mu^+ e^-$ scattering affected by $u$-channel $k_3$ exchange which interferes with the SM $Z/\gamma$ t-channel exchange. The resulting differential cross-section is\footnote{We use the leading order SM result in this analysis. See \cite{Broggio:2022htr} for the result at next-to-next-to-leading order.}
\begin{equation}
\begin{aligned}
&\frac{d\sigma}{du} = 
\frac{h_1^4 u^2}{64\pi s^2(u-m_{k_3}^2)^2}
+\frac{\alpha h_1^2}{4c_W^2s^2 } \frac{u^2(c_W^2 m_Z^2 -t)}{t(m_{k_3}^2-u)(m_Z^2-t)}
\\&+
\frac{\pi\alpha^2}{16s_W^4c_W^4} \frac{t^2(8s^2 s_W^4+u^2(1+16s_W^4))
-8 t m_Z^2 c_W^2 s_W^2(4s^2 s_W^2+u^2(1+4s_W^2))
+32 m_Z^4 c_W^4 s_W^4 (s^2+u^2)}{s^2 t^2(m_Z^2-t)^2}
\end{aligned}
\end{equation}
and the integration limits $u_+$ and $u_-$ in the laboratory frame are defined in Eqs.~\eqref{eq:intLimLabframe1} and \eqref{eq:intLimLabframe2}, where the minimum $u_-$ is taken when $\phi_{ee'} = 178^\circ$ and $u_+$ with $\phi_{\mu\tau'} = 2^\circ$. The integration limits sensitively depend on the energies of the electron and antimuon beams. Figure~\ref{fig:T3paramsp} shows the sensitivity to different beam energies. In the effective field theory limit we find for the cross section 
\begin{equation}
     \sigma \simeq 235\, \mathrm{pb}\left[1 -0.011\left(\frac{h_1 \mathrm{TeV}}{m_{k_3}}\right)^2+ 0.00032 \left(\frac{h_1 \mathrm{TeV}}{m_{k_3}}\right)^4 \right] 
 \end{equation}
using the beam energies $E_e=30$ GeV and $E_\mu=1$ TeV and the cuts described above. This results in the constraint on the Yukawa coupling
\begin{equation}
    h_1 \lesssim 0.17\, \frac{m_{k_3}}{\mathrm{TeV}}\; ,
\end{equation}
achievable after $1$ year of data taking.

 \section{Conclusion}
\label{sec:conc}

\begin{table}[htp!]
\begin{center}
\setlength{\tabcolsep}{10pt}
\renewcommand{\arraystretch}{1.5}
\begin{tabular}{|l|c|c|l|} \hline
Experiment & Process &  90\% C.L. limit & Assumptions  \\ \hline
Belle  & $\tau^-\to\mu^+e^-e^-$ & $\sqrt{f_1 f_2} \lesssim 0.17 \frac{m_{k_1}}{\mathrm{TeV}}$  & 782 fb$^{-1}$\\
Belle  & $\tau^-\to e^+\mu^-\mu^-$  &   $\sqrt{g_1 g_2} \lesssim 0.17 \frac{m_{k_2}}{\mathrm{TeV}}$  & 782 fb$^{-1}$\\
Belle II  & $\tau^-\to\mu^+e^-e^-$ &  $\sqrt{f_1 f_2} \lesssim 0.06 \frac{m_{k_1}}{\mathrm{TeV}}$  & 50 ab$^{-1}$ \\
Belle II  & $\tau^-\to e^+\mu^-\mu^-$ &   $ \sqrt{g_1 g_2} \lesssim 0.06 \frac{m_{k_2}}{\mathrm{TeV}}$  & 50 ab$^{-1}$  \\
DELPHI & $e^+e^- \to e^+e^-$ & $f_2\lesssim 1.4 \frac{m_{k_1}}{\mathrm{TeV}}$ & \\
DELPHI & $e^+e^- \to \mu^+\mu^-$ & $h_1\lesssim 0.72 \frac{m_{k_3}}{\mathrm{TeV}}$ & \\
DELPHI & $e^+e^- \to \tau^+\tau^-$ & $g_1\lesssim 0.66 \frac{m_{k_2}}{\mathrm{TeV}}$ & \\
\hline
$\mu^+ \mu^+$ collider  & $\mu^+\mu^+\to \tau^+e^+$   
& $\sqrt{g_1g_2}\lesssim 0.07\frac{m_{k_2}}{\rm TeV}$
 & 12 fb$^{-1}$, $\sqrt{s}=2$ TeV
\\
$\mu^+ \mu^+$ collider  &  $\mu^+\mu^+\to \mu^+\mu^+$  
& $g_2\lesssim 0.09\frac{m_{k_2}}{\rm TeV}$
& 12 fb$^{-1}$, $\sqrt{s}=2$ TeV
\\
$\mu^+e^-$ collider  & $\mu^+e^-\to e^+\tau^-$ 
&$\sqrt{f_1f_2}\lesssim 0.13\frac{m_{k_1}}{\rm TeV}$
& 100 fb$^{-1}$, $(E_e,E_\mu)=(30,1000)$ GeV
\\
$\mu^+e^-$ collider & $\mu^+e^-\to\tau^+\mu^-$ 
&$\sqrt{g_1g_2}\lesssim 0.13\frac{m_{k_2}}{\rm TeV}$
& 100 fb$^{-1}$, $(E_e,E_\mu)=(30,1000)$ GeV
 \\ 
$\mu^+e^-$ collider & $\mu^+ e^- \to \mu^+ e^-$ & $h_1\lesssim 0.17\frac{m_{k_3}}{\rm TeV} $ 
& 100 fb$^{-1}$, $(E_e,E_\mu)=(30,1000)$ GeV
\\ \hline
\end{tabular}
\end{center}
\caption{Summary table with main EFT limits for $\mu$TRISTAN processes compared with experimental constraints from Belle and DELPHI, and the Belle II sensitivity. The third column shows the 90\% C.L.\ upper bounds that either currently exist or can be placed on the parameter space of our models from different processes (second column) by different experiments (first column). The fourth column quotes the assumptions used to derive the limits. For Belle we show the current limit and for Belle II the one expected with 50~ab$^{-1}$ of data. For $\mu$TRISTAN, the bounds correspond to $1$ year of data taking and apply in the EFT limit, with the full non-EFT results for $\mu^+ \mu^+ \to e^+ \tau^+$ and $\mu^+ \mu^+ \to \mu^+ \mu^+$ given in Figs.~\ref{f:90CLmumu} and \ref{fig:sminterference} respectively. The limits for processes that do not occur in the SM are obtained using the Feldman-Cousins procedure \cite{Feldman:1997qc} assuming that no signal and background events are observed so that the experiment will place an upper limit of $N = 2.44$ events. For the SM-allowed processes $\mu^+\mu^+ \to \mu^+ \mu^+$ and $\mu^+e^- \to \mu^+e^-$, we quote the a 90\% C.L.\ limit that only takes into account the statistical sensitivity.
}
\label{tab:summary}
\end{table}%

A partial summary of our results is given in Table~\ref{tab:summary}, which lists the effective field theory limits, in the absence of resonance effects, achievable after 1 year of data taking using the default configurations of $\mu$TRISTAN compared to derived constraints from Belle and DELPHI, and the Belle II sensitivity. For $\mu^+ \mu^+ \to e^+ \tau^+$ and $\mu^+ \mu^+ \to \mu^+ \mu^+$, the full non-EFT results are depicted in Figs.~\ref{f:90CLmumu} and \ref{fig:sminterference} respectively. These plots also show the $s$-channel resonance effects that $\mu$TRISTAN would be sensitive to. Of course, these various results are not the ultimate $\mu$TRISTAN limits or sensitivities since several years of data taking would be expected, and it may be possible to run the accelerator at higher (or lower) centre-of-mass energies than the default setting. The effect of the latter is indicated in Figs.~\ref{f:90CLmumu}, \ref{f:90CLmue}, \ref{fig:sminterference} and \ref{fig:T3paramsp}.

We now summarise the competitiveness and complementarity of $\mu$TRISTAN with both existing results and prospects at Belle II:
\begin{itemize}\setlength{\itemsep}{0ex}
    \item $\mu^+ \mu^+ \to e^+ \tau^+$ (Fig.~\ref{f:90CLmumu}): $\mu$TRISTAN can exceed the parameter reach obtained by Belle, and after $1$ year to a few years of running would be competitive with, and complementary to, the ultimate reach of Belle II.
    \item $\mu^+ e^- \to e^+ \tau^-$ (Fig.~\ref{f:90CLmue}): Exceeds Belle, and would require several years of running to be comparable to the ultimate reach of Belle II.
    \item The case of $\mu^+ e^- \to \tau^+ \mu^-$ is similar to $\mu^+ e^- \to e^+ \tau^-$ with the replacements $f_{1,2} \to g_{1,2}$ and $m_{k_1} \to m_{k_2}$.
    \item The flavour-conserving and SM-allowed process $\mu^+ \mu^+ \to \mu^+ \mu^+$ can be uniquely probed by $\mu$TRISTAN for $s$-channel $k_2$ exchange effects (Fig.~\ref{fig:sminterference}).
    \item The flavour-conserving and SM-allowed process $\mu^+ e^- \to \mu^+ e^-$  can be probed for parameter-space regimes that far exceed the region disallowed by DELPHI (Fig.~\ref{fig:T3paramsp}).
\end{itemize}
We also emphasise that were Belle II to observe the relevant CLFV $\tau$ decays, then $\mu$TRISTAN could be used to study potential mediating particles more directly.

While we have framed our analysis in terms of simple lepton-triality models for the sake of concreteness, the general utility of $\mu$TRISTAN in both its $\mu^+ \mu^+$ and $\mu^+ e^-$ modes for exploring flavour-violating and flavour-conserving charged lepton scattering processes is evident. It would clearly provide complementary capability to Belle II's search for CLFV decay modes of the $\tau$ lepton.

\begin{acknowledgments}
We thank Geoff Taylor and Bruce Yabsley for useful discussions. This work was supported in part by Australian Research Council Discovery Project DP200101470. 
\end{acknowledgments}

\bibliography{refs}

\end{document}